\documentclass[a4paper,11pt]{article}
\usepackage{pos_2022}
\usepackage{multicol}
\usepackage{wrapfig}

\usepackage{graphicx}
\usepackage{slashed}
\usepackage{amsmath}
\usepackage{epsf}
\usepackage{epsfig}
\usepackage{epstopdf}
\usepackage{hyperref}
\usepackage{xcolor}
\usepackage{subcaption}
\usepackage{comment}
\usepackage{tikz}
\usetikzlibrary{positioning}

\newcommand{\be}{\begin{equation}}
\newcommand{\ee}{\end{equation}}
\newcommand{\bea}{\begin{eqnarray}}
\newcommand{\eea}{\end{eqnarray}}

\newcommand{\MSbar}{{\overline{\rm MS}}}
\newcommand{\GIRS}{{\rm GIRS}}
\newcommand{\LR}{{\rm L}}
\newcommand{\DR}{{\rm DR}}
\newcommand{\Tr}{{\rm Tr}}
\newcommand{\gtilde}{\frac{g^2}{16 \, \pi^2}\; }

\def\MSbar{\overline{\rm MS}}

\def\openone{\leavevmode\hbox{\small1\kern-6.8pt\normalsize1}}

\title{Supercurrent renormalization of $\cal{N}$ = 1 supersymmetric Yang-Mills theory on the lattice}
\ShortTitle{Supercurrent renormalization}

\author[a]{G. Bergner}
\author[b,c]{M. Costa}
\author[b]{H. Panagopoulos}
\author[d]{S. Piemonte}
\author*[a]{I. Soler}
\author[e]{G. Spanoudes}

\affiliation[a]{University of Jena, Institute for Theoretical Physics,\\
Max-Wien-Platz 1, 07743 Jena, Germany}

\affiliation[b]{Department of Physics, University of Cyprus,\\
1 Panepistimiou Avenue, 2109 Aglantzia, Nicosia, Cyprus}

\affiliation[c]{Department of Chemical Engineering, Cyprus University of Technology,\\
30 Archbishop Kyprianou Str., 3036, Limassol, Cyprus}

\affiliation[d]{University of Regensburg, Institute for Theoretical Physics,\\
Universit\"{a}tsstr. 31, 93040 Regensburg, Germany}

\affiliation[e]{Computation-based Science
and Technology Research Center,\\
The Cyprus Institute, 20 Kavafi Str., Nicosia 2121, Cyprus}

\emailAdd{georg.bergner@uni-jena.de}
\emailAdd{kosta.marios@ucy.ac.cy}
\emailAdd{panagopoulos.haris@ucy.ac.cy}
\emailAdd{stefano.piemonte@ur.de}
\emailAdd{ivan.soler.calero@uni-jena.de}
\emailAdd{g.spanoudis@cyi.ac.cy}

\abstract{Supersymmetry on the lattice is explicitly broken by the gluino mass and lattice artifacts. However, it can be restored in the continuum limit by fine tuning the parameters based on the renormalized Ward identities. On the renormalization step not only the mass but also the renormalization of the supercurrent needs to be addressed. Here we present a lattice investigation to obtain the renormalization factors of the supercurrent for $\mathcal{N}$=1 Super-Yang Mills theory in a gauge invariant renormalization scheme. We also provide the conversion factors which are necessary in order to translate our results to the more standard $\MSbar$ scheme.
\begin{center}
\includegraphics[scale=0.45]{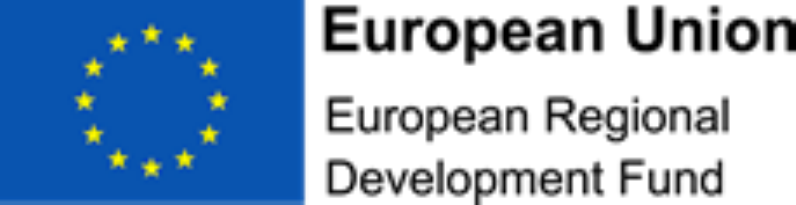}
\includegraphics[scale=0.45]{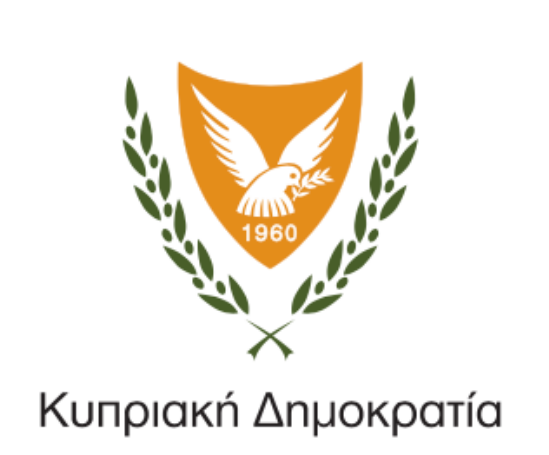}
\includegraphics[scale=0.45]{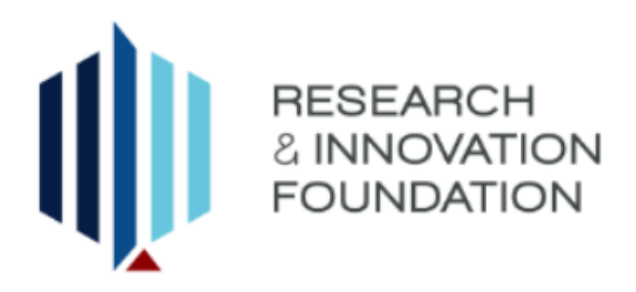}
\includegraphics[scale=0.45]{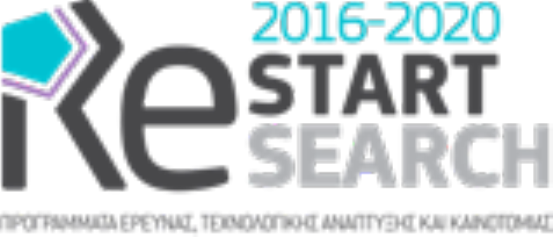}
\end{center}
}

\FullConference{%
The 39th International Symposium on Lattice Field Theory,\\
8th-13th August, 2022,\\
Rheinische Friedrich-Wilhelms-Universität Bonn, Bonn, Germany
}

\begin{document}
\maketitle

\section{Introduction}

Even though the lattice discretization breaks supersymmetry (SUSY)  explicitly, Curci and Veneziano \cite{Curci:1986sm} showed for $\mathcal{N} = 1 $ supersymmetric Yang-Mills (SYM)  that  chiral symmetry and supersymmetry can both be  recovered in the continuum limit by tuning a single parameter, the gluino mass. This approach has been successfully applied in numerical simulations providing also first insights about the particle spectrum of the theory \cite{Ali:2018fbq,Bergner:2015adz}.

In supersymmetric theories containing matter supermultiplets, such as SQCD, the number of parameters that need fine tuning is significantly larger. A tuning based on Ward identities becomes challenging. In particular renormalization coefficients of the supercurrent need to be determined numerically in this approach. In this work we study the renormalization of the supercurrent and compare non-perturbative effects to perturbation estimates.

As a first exploratory step we will study the renormalization of the supercurrent for $\mathcal{N} = 1 $ supersymmetric Yang-Mills (SYM) on the lattice. We will do so by computing, non-perturbatively, the renormalization factors in a gauge-invariant renormalization scheme (GIRS) \cite{Costa:2021iyv}. Furthermore, we will compute the conversion factors from GIRS to $\MSbar$ scheme perturbatively in dimensional regularization. Then we will be able to convert our lattice regularized GIRS renormalization factors to the more standard $\MSbar$ scheme.

\section{The model}
$\mathcal{N} = 1 $ SYM is the simplest  four-dimensional supersymmetric gauge theory. This theory describes the strong interactions between the carriers of the gauge force, the gluons, and their super-partners, the gluinos, which are Majorana fermions transforming under the adjoint representation of the gauge group. The gluons are represented by the non-Abelian gauge field $u^a_\mu(x)$ and the gluinos by the fermionic field $\lambda^a(x)$, where $a=1,...,N_c^{2}-1$. The on-shell Lagrangian for $\mathcal{N} = 1 $ in Euclidean space is
\begin{align}
    \mathcal{L}=\frac{1}{4}u^a_{\mu\nu}u^a_{\mu\nu} + \frac{1}{2}\bar{\lambda^a}\gamma_\mu (D_\mu \lambda)^a, \label{Lagrangian}
\end{align}
where $u^a_{\mu\nu}$ is the non-Abeliean field strength tensor and $D_\mu$ is the gauge covariant derivative acting as $(D_\mu\lambda)^a=\partial_\mu \lambda^a + g f_{abc}u_\mu^b\lambda^b$. The infinitesimal supersymmetry transformation leaving the action of the theory invariant is given by
\begin{align}
    &\delta u_\mu^a(x)=-i\bar{\xi}\gamma_\mu \lambda^a(x) \nonumber \\ 
    &\delta\lambda^a(x)=\frac{1}{2}\sigma_{\mu\nu}u_{\mu\nu}^a(x)\xi, \label{susy transf}
\end{align}
where $\sigma_{\mu\nu}=\frac{1}{2}[\gamma_\mu,\gamma_\nu]$ and $\xi$ is a Grassmann variable corresponding to the infinitesimal parameter of the transformation. Applying Noether's theorem to the classical theory, the symmetry transformation Eq.(\ref{susy transf}) leads to the conserved supercurrent
\be
S_\mu (x) \equiv - \sigma_{\nu \rho} \gamma_\mu {\rm tr}_c ( \,u_{\nu\,\rho}  (x) \lambda (x)).
\ee
Defining the theory at the quantum level requires regularization and renormalization which leads to important modifications. On the lattice, supersymmetry is broken by the addition of a gluino mass term and by the explicit breaking of Lorentz symmetry. Under renormalization the mass term gets additively renormalised and the supercurrent mixes with another dimension 7/2 operator
\be
T_\mu (x) \equiv 2\,\gamma_\nu {\rm tr}_c(\,u_{\mu\,\nu} (x) \lambda (x)). 
\ee
The corresponding Ward identity for the supercurrent after such modification reads as
\begin{align}
    Z_{SS}\big<\nabla_\mu S_\mu(x)Q(y)\big> + Z_{ST}\big<\nabla_\mu T_\mu(x)Q(y)\big>=m_S\big<\chi(x)Q(y)\big> + O(a),
\end{align}
where $Z_{SS}$ and $Z_{ST}$ are the renormalization coefficients of the supercurrent  $S^R_\mu=Z_{SS}S_\mu+Z_{ST}T_\mu$ and $m_s$ is the renormalized gluino mass; $Q(y)$ can be any operator localized at a point $y\ne x$.

We will explore in this work the renormalization of the Supercurrent operators $S_\mu, T_\mu$ on the lattice both numerically by Monte-Carlo simulations and using perturbation theory.

\section{Renormalization and GIRS scheme}
Our main goal is to compute the renormalization of the supercurrent both perturbatively and non-perturbatively. Therefore the first step is to decide on a proper renormalization scheme that can be applicable in both situations. In this work we will use the GIRS scheme ~\cite{Costa:2021iyv} which is reminiscent of the X-space renormalization scheme. The GIRS is defined through the renormalization conditions
\begin{align*}
				\langle\mathcal{O}^{\rm{GIRS}}_X(x)\mathcal{O}^{\rm{GIRS}}_Y(y)\rangle|_{_{x-y=\bar{z}}}\equiv
				Z_X^{\rm{GIRS}}Z_Y^{\rm{GIRS}}\langle\mathcal{O}^{\rm{B}}_X(x)\mathcal{O}^{\rm{B}}_Y(y)\rangle|_{_{x-y=\bar{z}}} = 
				\langle\mathcal{O}_X(x)\mathcal{O}_Y(y)\rangle^{\rm{tree}}|_{_{x-y=\bar{z}}}, 
\end{align*}
for $X, Y$ two operators of interest and $(x \neq y)$ in order to avoid potential contact terms; the superscript $B$ denotes bare quantities. This scheme is appealing because the two-point Green functions $\langle\mathcal{O}^{\rm{B}}_X(x)\mathcal{O}^{\rm{B}}_Y(y)\rangle|_{_{x-y=\bar{z}}}$ can be computed both non-perturbatively and also in perturbation theory. Even more important, choosing gauge-invariant operators, only the mixing between these operators and other gauge-invariant operators is relevant, which makes this scheme particularly suitable for lattice computations. Considering the gauge invariant operators $\mathcal{O}_X, \mathcal{O}_Y = S_\mu, T_\mu$ the resulting mixing matrix relating the bare and renormalized supercurrent operators is
\begin{gather}
 \begin{pmatrix} {S}^R_\mu & \\[2ex] {T}^R_{\mu} \end{pmatrix}
 =
    \begin{pmatrix}
   Z_{SS} &
   Z_{ST} \\[2ex]
   Z_{TS} &
   Z_{TT} 
   \end{pmatrix}
   \begin{pmatrix} {S}^B_{\mu} & \\[2ex] {T}^B_{\mu} \end{pmatrix}.
   \label{mixing matrix}
\end{gather}
To determine the 4 elements of the mixing matrix $Z$ we need 4 conditions:
\begin{itemize}
\item Three conditions can be imposed by considering expectation values between the two mixing operators \footnote{A bar on $S_\mu, T_\mu$ denotes the corresponding charge conjugates}
\begin{align}
G^{S\,S}_{\mu \nu}(x, y) \equiv \langle S_\mu (x) \ \overline{S}_\nu (y) \rangle\,\,\,&,\,\,\,
G^{T\,T}_{\mu \nu}(x, y) \equiv \langle T_\mu (x) \ \overline{ T}_\nu (y) \rangle\,\,\,,\,\,\, \nonumber \\[10pt]
G^{S\,T}_{\mu \nu}(x, y) \equiv &\langle S_\mu (x) \ \overline{ T}_\nu (y) \rangle.
\label{GFs}
\end{align}
\item A fourth condition can be imposed on two-point Green's functions involving products of $S_\mu$ (or $T_\mu$) with other gauge-invariant operators of equal or lower dimension. The only such operator with compatible behaviour under the Lorentz group is the Gluino-Glue operator ${\cal O} (x) \equiv \sigma_{\mu \nu} \,{\rm{tr}}_c (\,u_{\mu \nu} (x) \lambda (x))$ and a corresponding Green's function is
\begin{equation}
G^{{\cal O}\,S}_{\mu}(x, y) \equiv \langle {\cal O} (x) \ \overline{S}_\mu (y) \rangle.
\label{GFO}
\end{equation}
\end{itemize}
There is a variety of ways to imposed the GIRS renormalization conditions. Especially suitable for numerical lattice investigations is the following form where we integrate over the spatial components of $z = y - x = (\vec{z},t)$\, for the sake of improving the signal 
\begin{eqnarray}
\int d^3 \vec{z} \ {\rm Tr} \{ {\left[G^{S\,S}_{\mu \nu}(x, y)\right]}^{{\rm GIRS}} P_{\nu \mu} \} &=& \int d^3 \vec{z} \ {\rm Tr} \{ {\left[G^{S\,S}_{\mu \nu}(x, y)\right]}^{\rm tree} P_{\nu \mu} \}, \label{GIRS2_cond1} \\
\int d^3 \vec{z} \ {\rm Tr} \{ {\left[G^{T\,T}_{\mu \nu}(x, y)\right]}^{{\rm GIRS}} P_{\nu \mu} \} &=& \int d^3 \vec{z} \ {\rm Tr} \{ {\left[G^{T\,T}_{\mu \nu}(x, y)\right]}^{\rm tree} P_{\nu \mu} \}, \label{GIRS2_cond2} \\
\int d^3 \vec{z} \ {\rm Tr} \{ {\left[G^{S\,T}_{\mu \nu}(x, y)\right]}^{{\rm GIRS}} P_{\nu \mu} \} &=& \int d^3 \vec{z} \ {\rm Tr} \{ {\left[G^{S\,T}_{\mu \nu}(x, y)\right]}^{\rm tree} P_{\nu \mu} \}, \label{GIRS2_cond3} \\
\int d^3 \vec{z} \ {\rm Tr} \{ {\left[G^{S\,{\cal O}}_{\mu}(x, y)\right]}^{{\rm GIRS}} P_{\mu} \} &=& \int d^3 \vec{z} \ {\rm Tr} \{ {\left[G^{S\,{\cal O}}_{\mu}(x, y)\right]}^{\rm tree} P_{\mu} \}.
\label{GIRS2_cond4}
\end{eqnarray}
$P_{\nu \mu}=\gamma_\mu \gamma_4 \gamma_\nu$ and $P_{\mu}=\gamma_\mu \gamma_4$ are projectors acting on the Dirac space that project to states transforming properly under parity, time reversal and charge conjugation. The repeated indices $\mu, \nu$ are not summed over and there is a freedom on which components $\mu, \nu$ to choose. However, the operator components need to be the same in all GIRS conditions Eq.~(\ref{GIRS2_cond1}--\ref{GIRS2_cond4}), as in principle, in this scheme, different components could give different renormalization factors. The tree level values on the right hand side of Eq.~(\ref{GIRS2_cond1}--\ref{GIRS2_cond4}) after spatial integration are
\begin{align}
  \int d^3 \vec{z} \ {\rm Tr} \left[G^{SS, \,{\rm tree}}_{\mu \nu}(x, y) \ P_{\nu \mu} \right] &= - \frac{(N_c^2 - 1) \ t}{\pi^2 |t|^5} (1 - \delta_{\mu 4} - \delta_{\nu 4} - 3 \,\delta_{\mu \nu} + 4\, \delta_{\mu 4} \,\delta_{\nu 4}), \label{TL1GIRS2} \\
  \int d^3 \vec{z} \ {\rm Tr} \left[G^{TT, \,{\rm tree}}_{\mu \nu}(x, y) \ P_{\nu \mu} \right] &= \phantom{+} \frac{(N_c^2 - 1) \ t}{4 \pi^2 |t|^5} (2 + \delta_{\mu 4} + \delta_{\nu 4}  + 3 \,\delta_{\mu \nu} - 4 \,\delta_{\mu 4}\,\delta_{\nu 4}), \\
  \int d^3 \vec{z} \ {\rm Tr} \left[G^{ST, \,{\rm tree}}_{\mu \nu}(x, y) \ P_{\nu \mu} \right] &= - \frac{(N_c^2 - 1) \ t}{2 \pi^2 |t|^5} (1 - \delta_{\mu 4} - \delta_{\nu 4} - 3 \,\delta_{\mu \nu} + 4\, \delta_{\mu 4}\,\delta_{\nu 4}), \label{TL3GIRS2}\\
  \int d^3 \vec{z} \ {\rm Tr} \left[G^{S\mathcal{O}, \,{\rm tree}}_{\mu}(x, y) \ P_{\mu} \right] \ &= \phantom{+} 0. \label{TL4GIRS2}
\end{align}
It is instructive to write out the full set of GIRS conditions in terms of the bare correlators. They lead to a set of quadratic equations for the renormalization factors
{\small
\begin{align}
&Z_{SS}^2 \ {\rm Tr} \left[G^{SS}_{\mu \nu} P_{\nu \mu} \right] + Z_{SS} \ Z_{ST} \ ({\rm Tr} \left[G^{ST}_{\mu \nu} P_{\nu \mu} \right] + {\rm Tr} \left[G^{TS}_{\mu \nu} P_{\nu \mu} \right]) + Z_{ST}^2 \ {\rm Tr} \left[G^{TT}_{\mu \nu} P_{\nu \mu} \right] = {\rm Tr} \left[G^{SS,{\rm tree}}_{\mu \nu} P_{\nu \mu} \right], \nonumber\\
&Z_{TS}^2 \ {\rm Tr} \left[G^{SS}_{\mu \nu} P_{\nu \mu} \right] + Z_{TS} \ Z_{TT} \ ({\rm Tr} \left[G^{ST}_{\mu \nu} P_{\nu \mu} \right] + {\rm Tr} \left[G^{TS}_{\mu \nu} P_{\nu \mu} \right]) + Z_{TT}^2 \ {\rm Tr} \left[G^{TT}_{\mu \nu} P_{\nu \mu} \right] = {\rm Tr} \left[G^{TT,{\rm tree}}_{\mu \nu} P_{\nu \mu} \right], \nonumber\\
&Z_{SS} \ \left(Z_{TS} \ {\rm Tr} \left[G^{SS}_{\mu \nu} P_{\nu \mu} \right] + Z_{TT} \ {\rm Tr} \left[G^{ST}_{\mu \nu} P_{\nu \mu} \right]\right) + Z_{ST} \ \left(Z_{TS} \ {\rm Tr} \left[G^{TS}_{\mu \nu} P_{\nu \mu} \right] + Z_{TT} \ {\rm Tr} \left[G^{TT}_{\mu \nu} P_{\nu \mu} \right]\right) = \nonumber \\
&\hspace{12cm} {\rm Tr} \left[G^{ST,{\rm tree}}_{\mu \nu} P_{\nu \mu} \right], \nonumber \\
&Z_O \left(Z_{SS} \ {\rm Tr} \left[G^{SO}_{\mu} P_{\mu} \right] + Z_{ST} \ {\rm Tr} \left[G^{TO}_{\mu} P_{\mu} \right]\right) = {\rm Tr} \left[G^{SO,{\rm tree}}_{\mu} P_{\mu} \right] = 0, \label{cond_explicit}
\end{align}}where on the last equation we used the renormalization of the Gluino-Glue operator $\mathcal{O}^R=Z_\mathcal{O}\mathcal{O}^B$ and Eq.~(\ref{TL4GIRS2}). From Eq.~(\ref{TL1GIRS2}--\ref{TL3GIRS2})  one can see that the particular choice of temporal indices for either $\mu$ or $\nu$ will lead to a vanishing tree level value on the first equation in (\ref{cond_explicit}). This in combination with the last equation above would result in vanishing or indeterminate value for $Z_{SS}$ or $Z_{ST}$. Therefore the only allowed components $\mu$, $\nu$ are the spatial ones. 

\section{Perturbative results in dimensional regularization and conversion factors to \texorpdfstring{$\overline{\rm MS}$}{}}
\begin{figure}[ht]
\centering 
\includegraphics[width=.09\textwidth]{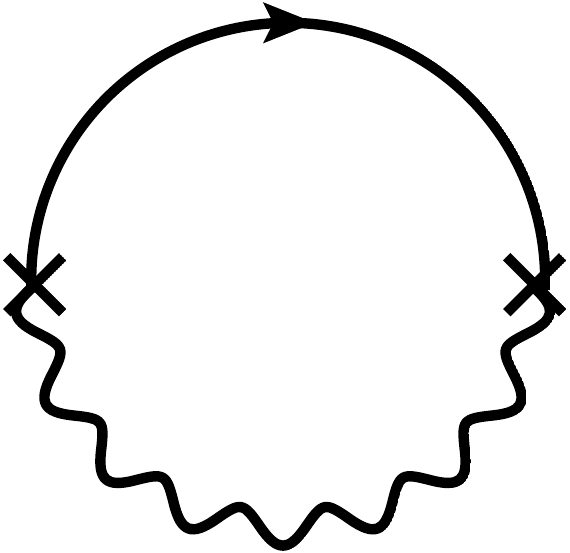}
\hfil
\includegraphics[width=.45\textwidth]{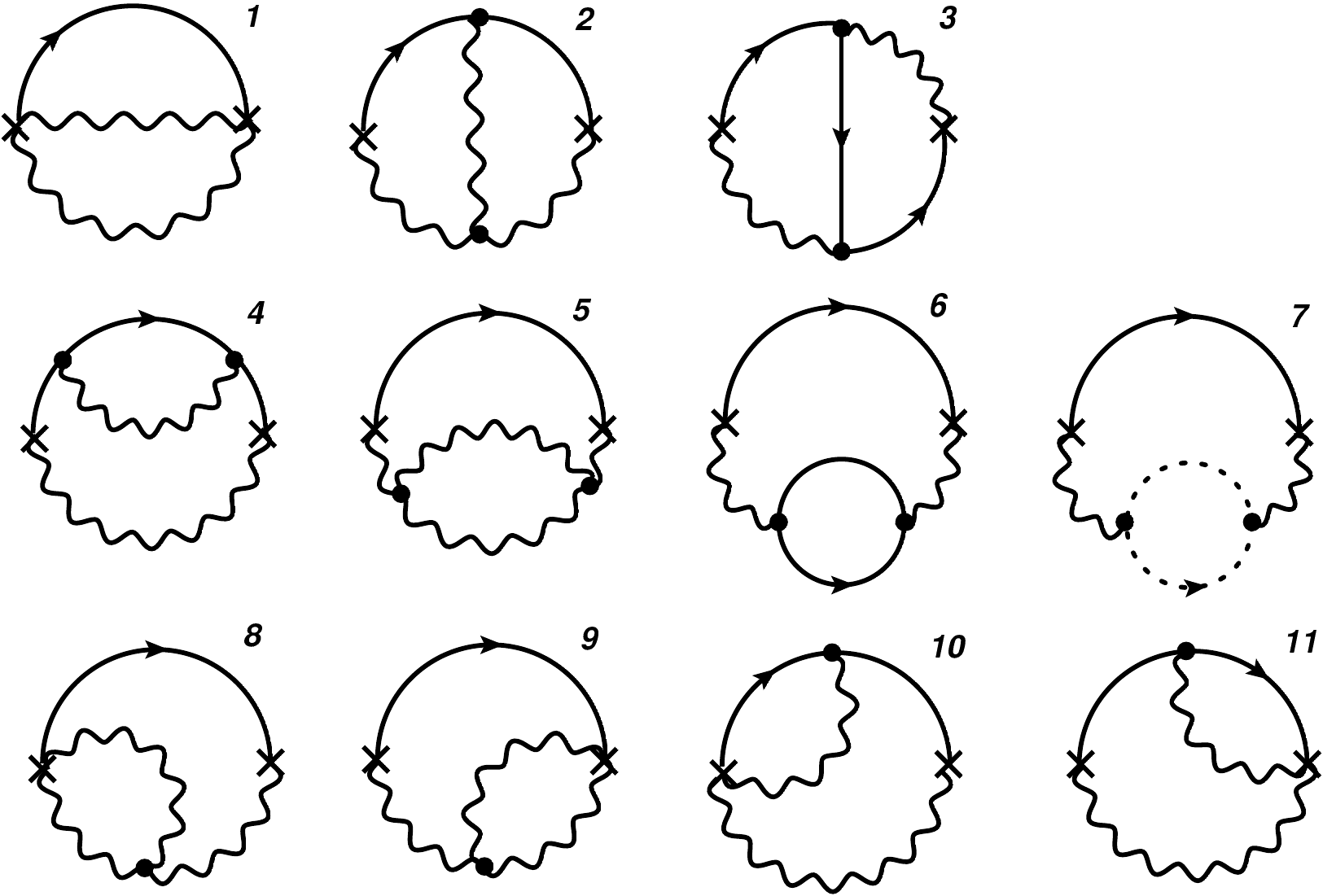}
\caption{One-loop and two-loop Feynman diagrams contributing to the tree-level and one-loop two-point Green's functions of Eqs.~(\ref{GFs}) and (\ref{GFO}). A wavy (solid, dashed) line represents gluons (gluinos, ghosts). The two crosses denote the insertions of operators $S_\mu,T_\nu, \mathcal{O}$ appearing in the definition of each two-point function.}
\label{diagrams}
\end{figure}

Instead of comparing perturbative and non-perturbative results on the GIRS scheme we will follow a different approach. We will convert our results from GIRS scheme to $\MSbar$ using the conversion factors $C^{GIRS, \MSbar}$. This comes with the advantage that $\MSbar$ is more amenable to perturbation theory and the fitting of the $Z$ factors on the numerical side is easier (see next section). The conversion factors $C^{GIRS, \MSbar}$ are the factors relating $\MSbar$ renormalized and GIRS renormalized operators
\begin{align}
		\begin{pmatrix}
			C_{SS}^{{\rm GIRS}, \ \overline{\rm MS}} & C_{ST}^{{\rm GIRS}, \ \overline{\rm MS}} \\
			\\
			C_{TS}^{{\rm GIRS}, \ \overline{\rm MS}} & C_{TT}^{{\rm GIRS}, \ \overline{\rm MS}}
		\end{pmatrix}  \cdot \begin{pmatrix}
		Z_{SS}^{{\ \rm R}, \ {\rm GIRS}} & Z_{ST}^{{\ \rm R}, \ {\rm GIRS}} \\
		\\
		Z_{TS}^{{\ \rm R}, \ {\rm GIRS}} & Z_{TT}^{{\ \rm R}, \ {\rm GIRS}}
		\end{pmatrix} & &=& &  \begin{pmatrix}
			Z_{SS}^{{\ \rm R}, \ \overline{\rm MS}} & Z_{ST}^{{\rm R }, \ \overline{\rm MS}} \\
			\\
			Z_{TS}^{{\ \rm R}, \ \overline{\rm MS}} & Z_{TT}^{{\rm R }, \ \overline{\rm MS}}
		\end{pmatrix}, \label{conversion_factors}
\end{align}
where R stands for a chosen regularization scheme and the conversion factors themselves are regularization independent. We first obtained the conversion factors by computing the renormalization constants to one-loop in perturbation theory in dimensional regularization both for the $\MSbar$ and the GIRS scheme. From the action defined by the Lagrangian Eq. (\ref{Lagrangian}), after applying the conventional Faddev-Popov method, one can obtain the corresponding bare Green's functions at tree-level and to one-loop order in dimensional regularization (DR), involving respectively, the one-loop and two-loop Feynman diagrams of Fig.~\ref{diagrams}.

As is standard practice, the pole terms ($1/\varepsilon^n$, $n \in \mathbb{Z}^+$) are removed by defining the $\MSbar$ mixing matrix elements to have only negative integer powers of $\varepsilon$, i.e., $Z_{ij}^{\DR,\MSbar} = \delta_{ij} + g^2 (z_{ij} / \varepsilon) + \mathcal{O} (g^4)$, where $i, j = S, T$ and $Z_{\mathcal{O}}^{\DR,\MSbar} = 1 + g^2 (z_{\mathcal{O}} / \varepsilon) + \mathcal{O} (g^4)$. Our results for $Z_{ij}^{\DR,\MSbar}$, $Z_{\mathcal{O}}^{\DR,\MSbar}$ read
\bea
Z_{SS}^{\DR,\MSbar} &=& 1 + \mathcal{O} (g^4), \label{ZSSMSbar} \\
Z_{ST}^{\DR,\MSbar} &=& \mathcal{O} (g^4), \\
Z_{TS}^{\DR,\MSbar} &=& \gtilde \frac{3 N_c}{2 \varepsilon} + \mathcal{O} (g^4), \\
Z_{TT}^{\DR,\MSbar} &=& 1 - \gtilde \frac{3 N_c}{\varepsilon} + \mathcal{O} (g^4), \label{ZTTMSbar} \\
Z_{\mathcal{O}}^{\DR,\MSbar} &=& 1 - \gtilde \frac{3 N_c}{\varepsilon} + \mathcal{O} (g^4),
\eea
which agree with our recent one-loop calculations in Refs.~\cite{Costa:2020keq, Bergner:2022wnb}. By combining our one-loop results for the mixing matrix in $\GIRS$ Eqs.~(\ref{cond_explicit}) and in $\MSbar$ Eqs.~(\ref{ZSSMSbar} -- \ref{ZTTMSbar}), we extract the one-loop conversion factors
\begin{align}
			C_{SS}^{{\rm GIRS},\ \overline{\rm MS}} &= 1 - \frac{g^2_{\overline{\rm MS}}}{16 \pi^2} \frac{17 N_c}{6} + \mathcal{O} (g^4_{\overline{\rm MS}}),\\
			C_{ST}^{{\rm GIRS},\ \overline{\rm MS}} &= \frac{g^2_{\overline{\rm MS}}}{16 \pi^2} 4 N_c + \mathcal{O} (g^4_{\overline{\rm MS}}), \\
			C_{TS}^{{\rm GIRS},\ \overline{\rm MS}}&= -\frac{g^2_{\overline{\rm MS}}}{16 \pi^2} \frac{3 N_c }{2} \left(\frac{2}{3} + 2 \gamma_E + \ln (\bar{\mu}^2 a \ t^2)\right) + \mathcal{O} (g^4_{\overline{\rm MS}}), \\
			C_{TT}^{{\rm GIRS},\overline{\rm MS}} &= 1 + \frac{g^2_{\overline{\rm MS}}}{16 \pi^2} N_c \left(\frac{7}{6} + 6 \gamma_E + 3 \ln (\bar{\mu}^2 a \ t^2)\right) + \mathcal{O} (g^4_{\overline{\rm MS}}).
		\end{align}

\section{Non-perturbative results}
For the lattice discretization of $\mathcal{N} = 1 $ SYM we employed: a tree-level Symanzik improved gauge action and Wilson fermions for the gluino fields. The action reads\footnote{ ${\rm tr}_c$ denotes trace over color matrices.}
\begin{align}
{\cal S}^{L}_{\rm SYM}=\sum_{x}\Bigg\{& \frac{2a^{4}}{g^2} \left[ \frac{5}{3} \sum_{\rm plaq.} {\rm Re} \ {\rm tr}_c (1- U_{\rm plaq.}) - \frac{1}{12} \sum_{\rm rect.} {\rm Re} \ {\rm tr}_c (1-U_{\rm rect.}) \right] + \sum_y\frac{a^3}{2\kappa}\bar{\lambda}(x)D_W\lambda(y)\Bigg\},\nonumber \\
\label{susylagrLattice}
\end{align}
where, $U_{\rm plaq.} (U_{\rm rect.})$ denotes $1 {\times} 1$ ($2 {\times} 1$) rectangular Wilson loops and the lattice Wilson operator is represented in terms of the hopping parameter $\kappa\equiv 1/(2m_0+8)$ as
\begin{equation}
    D_W = 1 - \kappa \big[(1-\gamma_\mu)(V_\mu(x))\delta_{x+\mu,y} + (1+\gamma_\mu)(V^\dagger{}_\mu(x-\mu))\delta_{x-\mu,y}\big].
    \label{eq:Dirac_Wilson}
\end{equation}
One-level of stout smearing was used on the links $V_\mu(x)$, which in the adjoint representation are given by $V_\mu^{ab}=2\,{\rm tr_c}[U^{\dagger}_\mu (x)T^a U_\mu(x)T^b]$.

We considered the configurations from earlier works \cite{Bergner:2015adz,Ali:2019agk} with ensembles based on two different gauge groups $SU(2)$ and $SU(3)$. In case of the gauge group $SU(3)$ the lattice action is different, see \cite{Ali:2019agk} for details. The ensembles were generated at two different lattice sizes $V=L^3\times T$ of $V_1=24^3\times48$ and $V_2=32^3\times64$ along with different values of the coupling constant $\beta$ and different mass parameters $\kappa$. As the behaviour of the renormalization constant did not change qualitatively from the different ensembles, we will present here the ones based on the $SU(2)$ group with a lattice size of $V_1=24^3\times48$, two different mass parameters $\kappa=0.14925,\ 0.14920$ and a gauge coupling of $\beta=1.75$. The complete set of results for all ensembles are collected in \cite{Bergner:2022see}.  For comparison of the results, the lattice spacing can be estimated using the QCD Sommer scale value $r_0=0.5\text{ fm}$ which leads to a lattice spacing of $a=0.0554(11) \text{ fm}$.

\begin{figure}[h]
\begin{minipage}{0.5\textwidth}
\begin{tikzpicture}
  \node (img)  {\includegraphics[width=.75\textwidth]{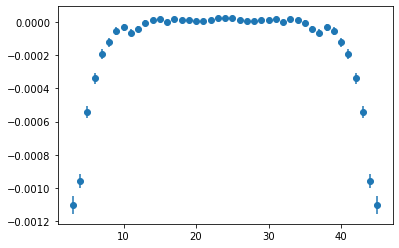}};
  \node[below=of img, node distance=0cm, xshift=0.5cm, yshift=1cm] {t};
  \node[left=of img, node distance=0cm, rotate=90, anchor=center,yshift=-0.7cm,] {$\text{Tr}\langle\mathcal{O}(t)S_i(0)P_i\rangle$};
 \end{tikzpicture}
\end{minipage}%
\begin{minipage}{0.5\textwidth}
\begin{tikzpicture}
  \node (img)  {\includegraphics[width=.75\textwidth]{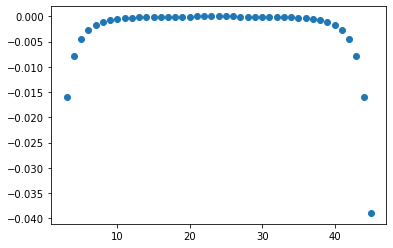}};
  \node[below=of img, node distance=0cm, xshift=0.5cm, yshift=1cm] {t};
  \node[left=of img, node distance=0cm, rotate=90, anchor=center,yshift=-0.7cm,] {$\text{Tr}\langle\mathcal{O}(t)T_i(0)P_i\rangle$};
 \end{tikzpicture}
\end{minipage}%
\caption{Correlators $\text{Tr}\langle\mathcal{O}(t)S_i(0)P_i\rangle$ and $\text{Tr}\langle\mathcal{O}(t)T_i(0)P_i\rangle$ computed numerically on the ensemble with $\kappa=0.14925$ and $\beta=1.75$ of the $V=24^3\times48$ lattice.}
\label{fig:correlators}
\end{figure}
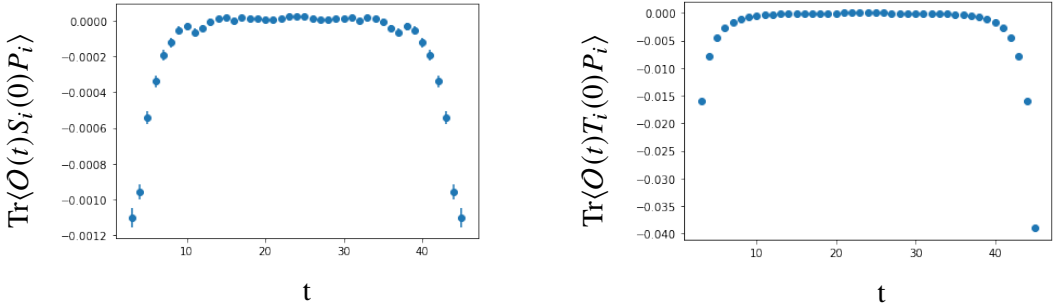

The supercurrent and $\mathcal{O}$ operators are represented on the lattice using clover plaquettes $\hat{F}_{\mu\nu}^{\alpha\beta}(x,t)$ and gluino fields $\lambda(x)$. 
The correlators between $S_\mu$, $T_\mu$, and $\mathcal{O}$  take the following generic omitting Lorentz, spinor and color indices
\begin{equation}
   \langle A(t)\overline{B}(0)\rangle \equiv \sum_{\vec{x},\vec{y}}\langle A(\vec{x},t)\overline{B}(\vec{y},0)\rangle = \sum_{\vec{x},\vec{y}} \langle\Tr[\Gamma \hat{F}(\vec{x},t)D^ {-1}(\vec{x},t|\vec{y},0)\hat{F}(\vec{y},0)\Gamma']\rangle_G = C^{\alpha \beta},
\end{equation}
for $A, B=S_\mu,T_\mu,\mathcal{O}$. The expectation value $\langle \cdot \rangle_G$ indicates that the fermion has been integrated out. $\Gamma$ and $\Gamma'$ collect the combination of gamma matrices of each operator and the inverse of the Dirac operator  $D^{-1}(x|y)$ propagates a gluino from $x$ to $y$.
In order to use the set of four GIRS conditions Eq.~(\ref{GIRS2_cond1}--\ref{GIRS2_cond4}) we are constrained to use spatial projectors $P_i=\gamma_4\gamma_i$ and $P_{ij}=\gamma_i \gamma_4 \gamma_j$ with $i=1,2,3$. We chose $i=j$ which has the advantage of giving better signal to noise ratio. As an example we present two of the correlators in Fig.~\ref{fig:correlators}.

Due to the gauge nature of these operators one is expected to find a substantial amount of noise on the signal. To improve the signal we used isotropy, time reversal, and charge conjugation to average equivalent correlators. We used wall sources for the operators and we average over spatial positions of the sink. This amounts to summing up over all $\vec{x}$ and $\vec{y}$ contributions of the correlator. Using the GIRS conditions which on the lattice take the following form
\bea
\frac{1}{3 L^3} \sum_{\vec{x},\vec{y}} \sum_i \ {\rm Tr} \left[G^{SS, \,\GIRS}_{ii}((\vec{x},t), (\vec{y},0)) \ \gamma_i\gamma_4\gamma_i \right] &=&  \frac{2 (N_c^2 - 1) t}{\pi^2 {|t|}^5}, \label{lattGIRS_cond1} \\
\frac{1}{3 L^3} \sum_{\vec{x},\vec{y}} \sum_i \ {\rm Tr} \left[G^{TT, \,\GIRS}_{ii}((\vec{x},t), (\vec{y},0)) \ \gamma_i\gamma_4\gamma_i \right] &=&  \frac{5 (N_c^2 - 1) t}{4 \pi^2 {|t|}^5}, \label{lattGIRS_cond2} \\
\frac{1}{3 L^3} \sum_{\vec{x},\vec{y}} \sum_i \ {\rm Tr} \left[G^{ST, \,\GIRS}_{ii}((\vec{x},t), (\vec{y},0)) \ \gamma_i\gamma_4\gamma_i \right] &=&  \frac{(N_c^2 - 1) t}{\pi^2 {|t|}^5}, \label{lattGIRS_cond3} \\
\frac{1}{3 L^3} \sum_{\vec{x},\vec{y}} \sum_i \ {\rm Tr} \left[G^{S{\cal O}, \,\GIRS}_{i}((\vec{x},t), (\vec{y},0)) \ \gamma_4 \gamma_i \right] &=& 0. \label{lattGIRS_cond4} 
\eea

Again these conditions lead to a set of second order equations for the renormalization factors $Z$ similar to Eq.~\eqref{cond_explicit} but in their lattice discretized versions.

In GIRS the time separation $t$ represents an energy scale for the renormalization constants. The short distance part is dominated by contact terms and lattice artifacts and has to be neglected. We use the conversion factors explained in previous sections to convert GIRS to $\MSbar$ scheme. This is expected to replace the dependence on the GIRS scale with the  the one on the $\overline{\mu}$ energy scale of the $\MSbar$ scheme. After the conversion, a plateau like behaviour is expected at larger distances and the dependence on time separation is replaced by a dependence on the energy scale. More importantly, converting to the $\MSbar$ scheme will allow us to compare directly with other results in perturbation theory.

\begin{figure}[ht]
    \centering
    \begin{subfigure}{1\textwidth}
    \begin{tikzpicture}
    \node (img)  {\includegraphics[width=.45\textwidth]{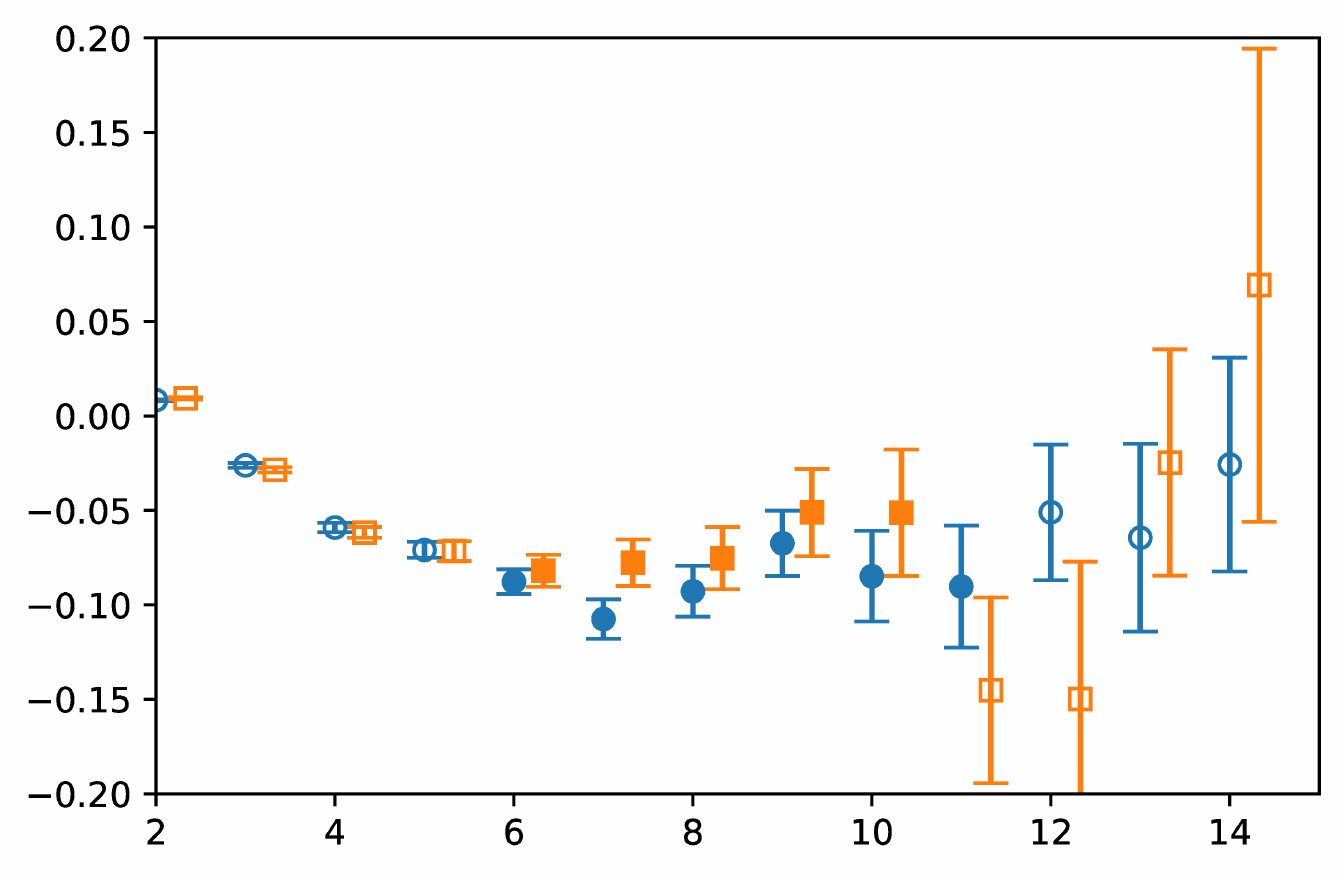}};
      \node[above,font=\bfseries] at (current bounding box.north) {GIRS scheme};
      \node[below=of img, node distance=0cm, xshift=0.5cm, yshift=1cm] {t};
      \node[left=of img, node distance=0cm, rotate=90, anchor=center,yshift=-0.7cm,] {$Z_{ST}/Z_{SS}$}; 
     \end{tikzpicture}
    \begin{tikzpicture}
    \node (img)  {\includegraphics[width=.45\textwidth]{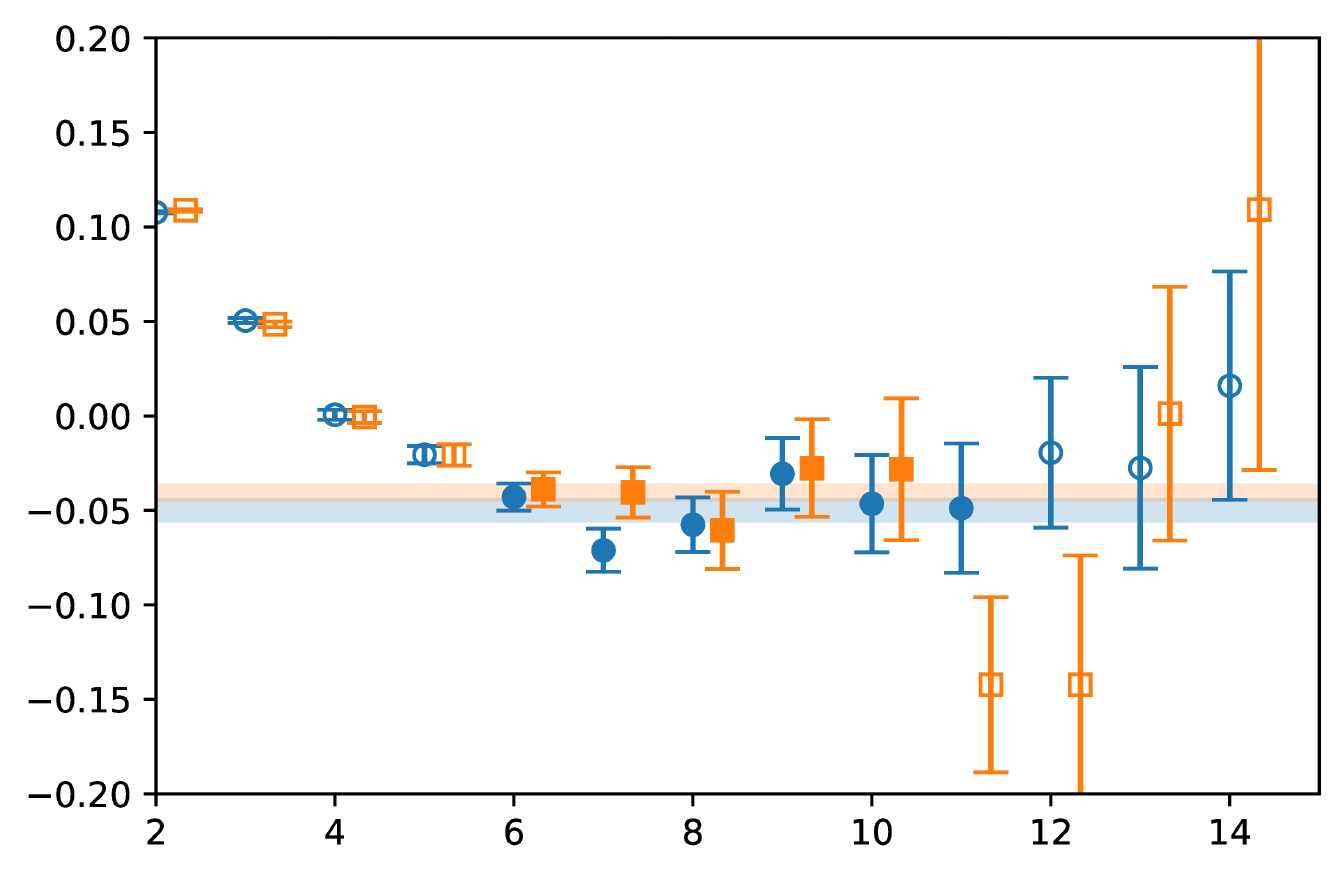}};
      \node[above,font=\bfseries] at (current bounding box.north) {$\MSbar$ scheme};
      \node[below=of img, node distance=0cm, xshift=0.5cm, yshift=1cm] {t}; 
      \end{tikzpicture}
    \centering
    \end{subfigure}
\caption{ $V=24^3\times 48$ lattice with $\kappa=0.14925$ (blue dots) and $\kappa=0.14920$ (orange squares). The points on the $\kappa=0.14920$ ensemble are shifted in $t$ by $+0.33$ for visibility. The error bars were estimated by a jackknife analysis and measurements were done every 8th step on Monte-Carlo time to reduce autocorrelations.}
\label{fig:Z_factor}
\end{figure}
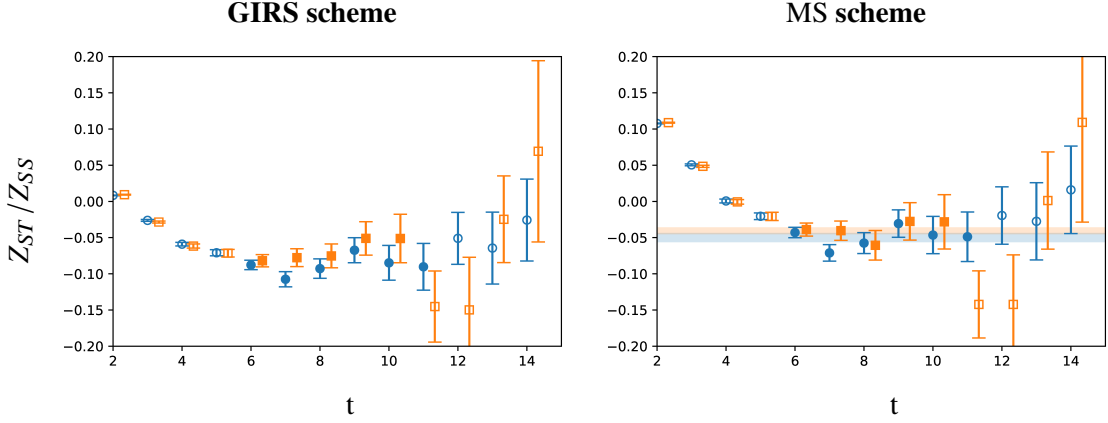

The result after applying the conversion factors to the $Z_{ST}/Z_{SS}$ factor is shown in Fig.~\ref{fig:Z_factor}. We fitted the data points in time interval $t\in[5,11]$ where contact terms have decayed and the noise has still not overcome the signal. The value obtained is $Z_{ST}/Z_{SS}=-0.0418(84)$ while the perturbative one found in \cite{Bergner:2022wnb} but without clover improvement is $Z_{ST}/Z_{SS}=0.10080$. 

\section{Summary and conclusions}
We studied the renormalization of the supercurrent for $\mathcal{N} = 1 $ SYM. We extracted the renormalization factors for the $S_\mu$ and $T_\mu$ operators in the GIRS scheme from bare correlators computed numerically on the lattice. By finding the conversion factors $C^{GIRS, \MSbar}$ perturbatively, we translated the non-perturbative results to the $\MSbar$ scheme where we could compare to perturbation theory.

We observed a significant disagreement between the perturbative and non-perturbative determination of the $Z$ factors. Yet we have room for improvement: simulating closer to the perturbative regime, including two-loop terms in the perturbative computation or including smearing could improve the agreement. They are feasible albeit complicated tasks, without any conceptual hindrances.

It is worth reiterating at this point that the determination of $Z_{ST}^{\LR,\MSbar} / Z_{SS}^{\LR,\MSbar}$ via GIRS, despite the present discrepancy with perturbative estimates, stands to be very useful in the study of more complicated theories, such as supersymmetric QCD, for the purpose of reducing the number of undetermined parameters in Ward identities.

\noindent{\bf Acknowledgements:} M.C., H.P. and G.S. acknowledge financial support from the European Regional Development Fund and the Cyprus Research and Innovation Foundation (Projects: EXCELLENCE/0918/0066 and EXCELLENCE/0421/0025. M.C. also acknowledges partial support from the Cyprus University of Technology under the "POSTDOCTORAL" program. G.S acknowledges financial support from H2020 project PRACE-6IP (Grant agreement ID: 823767). G.B. and I.S. acknowledge financial support from the Deutsche Forschungsgemeinschaft (DFG) Grant No.~432299911 and 431842497

\begin{thebibliography}{99}
\bibitem{Curci:1986sm}
G.~Curci and G.~Veneziano,
\emph{Supersymmetry and the Lattice: A Reconciliation?},
\href{https://doi.org/10.1016/0550-3213(87)90660-2}
{\emph{Nucl. Phys. B} \textbf{292} (1987) 555-572}.

\bibitem{Ali:2018fbq}
S.~Ali et al.,
\emph{Analysis of Ward identities in supersymmetric Yang-Mills theory},
\href{https://doi.org/10.1140/epjc/s10052-018-5887-9}
{\emph{Eur. Phys. J. C} \textbf{78} (2018) 404}
[\href{https://arxiv.org/abs/1802.07067}
{hep-lat/1802.07067}].

\bibitem{Bergner:2015adz}
G.~Bergner, P.~Giudice, G.~M\"unster, I.~Montvay and S.~Piemonte, 
\emph{The light bound states of supersymmetric $SU(2)$ Yang-Mills theory},
\href{https://doi.org/10.1007/JHEP03(2016)080}
{\emph{JHEP} \textbf{03} (2016) 080}
[\href{https://arxiv.org/abs/1512.07014}
{hep-lat/1512.07014}].

\bibitem{Costa:2021iyv}
M.~Costa et al.,
\emph{Gauge-invariant renormalization scheme in QCD: Application to fermion bilinears and the energy-momentum tensor},
\href{https://doi.org/10.1103/PhysRevD.103.094509}
{\emph{Phys. Rev. D} \textbf{103} (2021) 094509}
[\href{https://arxiv.org/abs/2102.00858}
 {hep-lat/2102.00858}].

\bibitem{Bergner:2022wnb}
G.~Bergner, M.~Costa, H.~Panagopoulos, I.~Soler and G.~Spanoudes,
\emph{Perturbative renormalization of the supercurrent operator in lattice N=1 supersymmetric Yang-Mills theory},
\href{doi:10.1103/PhysRevD.106.034502}
{\emph{Phys. Rev. D} \textbf{106} (2022) 034502}
[\href{https://arxiv.org/abs/2205.02012}
{hep-lat/2205.02012}].

\bibitem{Costa:2020keq}
M.~Costa, H.~Herodotou, P.~Philippides and H.~Panagopoulos,
\emph{Renormalization and mixing of the Gluino-Glue operator on the lattice},
\href{doi:10.1140/epjc/s10052-021-09173-x}
{\emph{Eur. Phys. J. C} \textbf{81} (2021), 401}
[\href{https://arxiv.org/abs/2010.02683}
{hep-lat/2010.02683}].

\bibitem{Ali:2019agk}
S.~Ali, G.~Bergner, H.~Gerber, I.~Montvay, G.~M\"unster, S.~Piemonte and P.~Scior,
\emph{Numerical results for the lightest bound states in $\mathcal{N}=1$ supersymmetric SU(3) Yang-Mills theory},
\href{doi:10.1103/PhysRevLett.122.221601}
{\emph{Phys. Rev. Lett.} \textbf{122} (2019) 221601}
[\href{https://arxiv.org/abs/1902.11127}
{hep-lat/1902.11127}].

\bibitem{Bergner:2022see}
G.~Bergner, M.~Costa, H.~Panagopoulos, S.~Piemonte, I.~Soler and G.~Spanoudes,
\emph{Nonperturbative renormalization of the supercurrent in $\mathcal{N} = 1$ Supersymmetric Yang-Mills Theory}, 
[\href{https://arxiv.org/abs/2209.13934}
{hep-lat/2209.13934}].

\end{thebibliography}
\end{document}